\documentclass{mn2e}
\usepackage{times}
\input{epsf}
\input{rotate}

\newcommand{\bb}{\bibitem[]{bla}}
\newcommand{\zm}{ \relax \ifmmode {\rm M_{\odot}} \else {M$_{\odot}$}\fi}

\newcommand{\ang}{$\rm \AA$}

\newcommand{\degree}{$^{\rm o}$}

\newcommand{\mic}{$\mu$m}
\newcommand{\ea}{{et al.}}

\newcommand{\km}{km s$^{-1}$}
\newcommand{\ha}{H$\alpha$}

\def\kms{\,km~s$^{-1}$}      

\def\lesssim{\mathrel{\hbox{\rlap{\hbox{\lower4pt\hbox{$\sim$}}}\hbox{$<$}}}}
\def\gtrsim{\mathrel{\hbox{\rlap{\hbox{\lower4pt\hbox{$\sim$}}}\hbox{$>$}}}}

\def\ion#1#2{#1$\;${\small\rm\@Roman{#2}}\relax}
 
\makeatletter 
\renewcommand\@biblabel[1]{}     

\makeatother

\begin{document}

\title[Pa$\beta$ spectropolarimetry]
{ 
Near-infrared line spectropolarimetry of
hot  massive stars }

\author[Ren\'e D. Oudmaijer \ea ]
{ Ren\'e  
D. Oudmaijer$^{1}$, Janet E. Drew$^{2}$, Jorick S. Vink$^2$ \\
$^{1}$
School of Physics \& Astronomy, EC Stoner Building, University of Leeds, 
 Leeds LS2 9JT, U.K.\\
$^{2}$
Imperial College of Science, Technology and Medicine,
Blackett Laboratory, Prince Consort Road, London,  SW7 2BZ, U.K.   \\
}

\date{received,  accepted}

\maketitle
\begin{abstract}
In order to study the inner parts of the circumstellar material around
optically faint, infrared bright objects, we present the first
medium-resolution spectropolarimetric data taken in the near-infrared.
In this paper we discuss Pa$\beta$ line data of GL 490, a well-known
embedded massive young stellar object, and of MWC 349A and MWC 342,
two optically faint stars that are proposed to be in the pre-main
sequence phase of evolution.  As a check on the method, the classical
Be star $\zeta$ Tau, known to display line polarization changes at
optical wavelengths, was observed as well.  Three of our targets show
a ``line effect'' across Pa$\beta$.  For $\zeta$ Tau and MWC 349A this
line effect is due to depolarisation by a circumstellar
electron-scattering disk. In both cases, the position angle of the
polarisation is consistent with that of the larger scale disks imaged
at other wavelengths, validating infrared spectropolarimetry as a
means to detect flattening on small scales. The tentative detection of
a rotation in the polarization position angle at Pa$\beta$ in the
embedded massive young stellar object GL 490 suggests the presence of
a small scale rotating accretion disk { with an inner hole} --
similar to those recently discovered at optical wavelengths in Herbig
Ae and T Tauri stars.

\end{abstract}


\section{Introduction}

Many open questions regarding the formation and evolution of stars
require the ability to probe the inner parts of their circumstellar
material, where accretion processes occur and stellar outflows find
their origin.  A powerful tool to address such issues is spectral line
polarimetry.  The method can utilise the fact that free electrons in
ionised material scatter photons and thereby polarise photospheric
radiation. In the case of spherical symmetry, all polarisation vectors
cancel, and if the envelope is unresolved a net zero polarisation is
observed. On the other hand, if the circumstellar geometry is
aspherical, such as a disk, a measurable net polarisation can be
observed.  This aspect of the method exploits the fact that ionised
gas also emits recombination lines, which, by virtue of their location
within the extended ionised gas undergo less scatterings by electrons,
and will thus be less polarised. A ``line-effect'' is then observed.
Electron scattering typically results in polarisation of order 1\%
(see Poeckert \& Marlborough 1976) and occurs at scales of order
several stellar radii (e.g. Cassinelli, Nordsieck \& Murison, 1987).
The method can be used for objects that display emission lines, and
immediately provides the answer to the question whether the
circumstellar (electron-scattering) material is spherically symmetric
when the source itself cannot be resolved in imaging (e.g. Oudmaijer
\& Drew 1999, Harries et al. 2002, Schulte-Ladbeck et al. 1993).  A
further advantage of spectropolarimetry is that it provides additional
constraints to straightforward spectroscopy, which is particularly
relevant for follow-up modelling (Drew et al. 2004, Harries 2000,
Vink, Harries \& Drew 2005a).

Given the debate in the massive star formation community about the
respective roles of isolated disk accretion versus competitive
accretion in dense cluster environments or even mergers of lower mass
protostars, it is crucial to add discriminating observational
constraints to this discussion. In particular, it has been argued that
the absence of accretion disks in massive young stellar objects (YSOs)
is reason to question the relevance of the disk accretion scenario
(e.g. Wolfire \& Cassinelli 1987; see also Norberg \& Maeder
2000). Although a positive detection of a disk around a YSO should by
no means be seen as proof for disk accretion, a continuing absence of
reliable disk signatures may eventually be taken as evidence against
it.

We have obtained spectropolarimetric data of the intermediate mass
pre-main sequence Herbig Ae/Be stars using the H$\alpha$ emission line
(Oudmaijer \& Drew, 1999, Vink et al. 2002).  More than half of the
Herbig Ae/Be stars show the line-effect, while the position angles of
the circumstellar material are consistent with the, larger scale,
disks that in some cases have been imaged by other techniques (see the
compilation by Vink et al. 2005b). The high proportion of line effect
detections in these objects suggests that {\it all} Herbig Ae/Be stars
are surrounded by disks on small scales. This result provides indirect
support to the notion that Herbig Ae/Be stars have grown via
accretion.

However,  more massive stars with even stronger radiation pressure
have remained elusive.  These most massive young stellar objects stay
embedded in their parental cloud until after settling on the zero-age
main sequence, which prevents the use of optical line polarimetry.  It
is therefore necessary to go to (near-)infrared wavelengths to study
such objects. This has never been done so far, presumably because
common-user polarisation optics at near-infrared optimised telescopes
are rare. To do such a study, we first need to consider which line to
observe, which is no trivial task. One has to take into account the
intrinsic strength of the line, the large extinction towards these
objects and the contribution of excess emission due to thermally
re-radiating dust. At the longer $K$-band wavelengths the sources
suffer less from extinction, but the dust excess emission may set an
upper limit to the observable line de-polarisations. This is because
the magnitude of the ``line-effect'' is determined by the strength of
the intrinsic continuum polarisation against which the change across
the unpolarized line is contrasted. Any excess emission due to an
extended envelope will effectively dilute the stellar continuum
polarisation and consequently this requires more sensitive data. As
these issues can not be answered easily pending dedicated
observations, we decided to conduct a pilot study to observe optically
faint objects and, for the first time, apply this method in the
near-infrared. We chose to use the Pa$\beta$ line in the $J$-band, one
of the strongest hydrogen recombination and at a wavelength where the
observable continuum radiation is more likely to be dominated by the
star.

We observed four objects, MWC 349A, MWC 342 and GL 490, and as a check
$\zeta$ Tau.  The first two objects are optically faint and are
proposed to be young pre-main sequence stars while GL 490 is a
well-known embedded massive young stellar object.  $\zeta$ Tau is a
classical Be star, known to display a line effect at optical
wavelengths. It was observed as well to test and prove the concept.

\begin{table}
\caption{Targets
\label{targets}}
\begin{tabular}{llllllll}
\hline
\hline
Name          &  {\it V }&{\it J} & t$_{\rm exp}$ (s) & $P_{\rm cont}$ (\%)  & $\Theta_{\rm cont}$ (\degree ) \\
\hline
\hline
GL 490        &  14  &10.5      & 4800 &  15.59$\pm$ 0.05  & 130.1  $\pm$ 0.3 \\
MWC 349A      &  13  & 7.0      &  800 &   3.60$\pm$ 0.04  & 176.6  $\pm$ 0.3 \\
$\zeta$ Tau   &  3   & 3.2      &  240 &   1.10$\pm$ 0.005 &  35.4  $\pm$ 0.1 \\
MWC 342       &  10.6 &$\sim$ 6 &  320  &  0.40$\pm$ 0.03  & 152.5  $\pm$ 2.3   \\

\hline
\hline
\end{tabular}
\, \\
\noindent

Continuum polarisations are measured in line-free regions in the
spectra. The error bars are internal errors, and small due to the large
number of pixels that are averaged. The external errors are estimated
to be 0.15\% and 2\degree .

\end{table}

\section{Observations \& Data reduction}

The observations were taken during the night of the 26$^{\rm th}$
September 1999 using the UKIRT telescope on Mauna Kea, Hawaii.  The
weather was reasonable, with a seeing of order 1.3 arcsec.  The CGS4
spectrograph was used in conjunction with the 150 lines/mm grating and
the long camera. The detector was a 256$\times$256 InSb array which has
a pixel size projected on the sky of 0.6 arcsec. During the
observations, the slitwidth was 4 pixels (2.4 arcsec), and the slit
was oriented East-West. A krypton arc-lamp was used for wavelength
calibration. This set-up resulted in a wavelength range of 1.256 -
1.31 $\mu$m, with pixels of $\sim$ 2.1 $\times 10^{-4} \mu$m.  The
final spectra have a spectral resolution of 170 \km . The objects do
not appear to be extended in our data.

For the linear polarisation observations, the IRPOL2 instrument was
employed. This polarimetry module consists of a MgF$_2$ prism and a
half-waveplate. The QU spectra were recorded at various positions of
the waveplate, while the objects were nodded up and down the slit to
facilitate sky-subtraction.  The observing sequence consisted of
consecutive observations of source and sky (by nodding over the slit)
for every position of the waveplate.  Polarised and un-polarised
standards were observed throughout the night.  A log of the
observations is provided in Table~\ref{targets}.

The data were reduced in {\sc iraf}, and included flatfielding, sky
and bias subtraction, bad pixel masking and extraction of the
spectra. The resulting spectra were then imported in the
Starlink\footnote{http://star-www.rl.ac.uk/} Time Series and
Polarimetry {\sc tsp} software, in which the Stokes parameters were
determined. Further manipulations were done in the Starlink {\sc
polmap} package for the polarisation spectra and {\sc iraf} for the
intensity spectra.

From the two unpolarised standards, HD~10476 and HD~202573, we
determined the instrumental polarisation to be 0.165\% at an angle of
122.2\degree . This value was subtracted from the reduced data. The
zero-point of the polarisation angle was determined assuming the
polarisation standard stars HD 23512 and HD 43384 to have polarisation
angles of 29.6\degree \ and 170.7\degree \ respectively. These values
correspond to the {\it R} band values listed by Hsu \& Breger
(1982). The corrections were within 1\degree \ from each other.  The
polarisation measured in the standard stars at 1.28 $\mu$m is slightly
(0.1\%) larger than an extrapolation from their optical data using a
Serkowski law would suggest. This is a well-documented effect, the
polarisation in the near-infrared drops of shallower than in the
optical (e.g. the review by Whittet, 1996).

The errors in polarisation per pixel range from $\sim 0.05-0.1\%$ for
the brighter stars to 0.25\% for MWC 342 and MWC 349A and 0.40\% for
GL 490. For the brightest objects, the error in the polarisation is
observed to increase with increasing wavelength. This is most likely
due to low level fringing occurring in the optics. The final,
external, accuracies are estimated to be of order 0.15\% and 2\degree
\ respectively.

\begin{figure*}
 \mbox{\epsfxsize=0.6\textwidth\epsfbox[1 600 300 800]{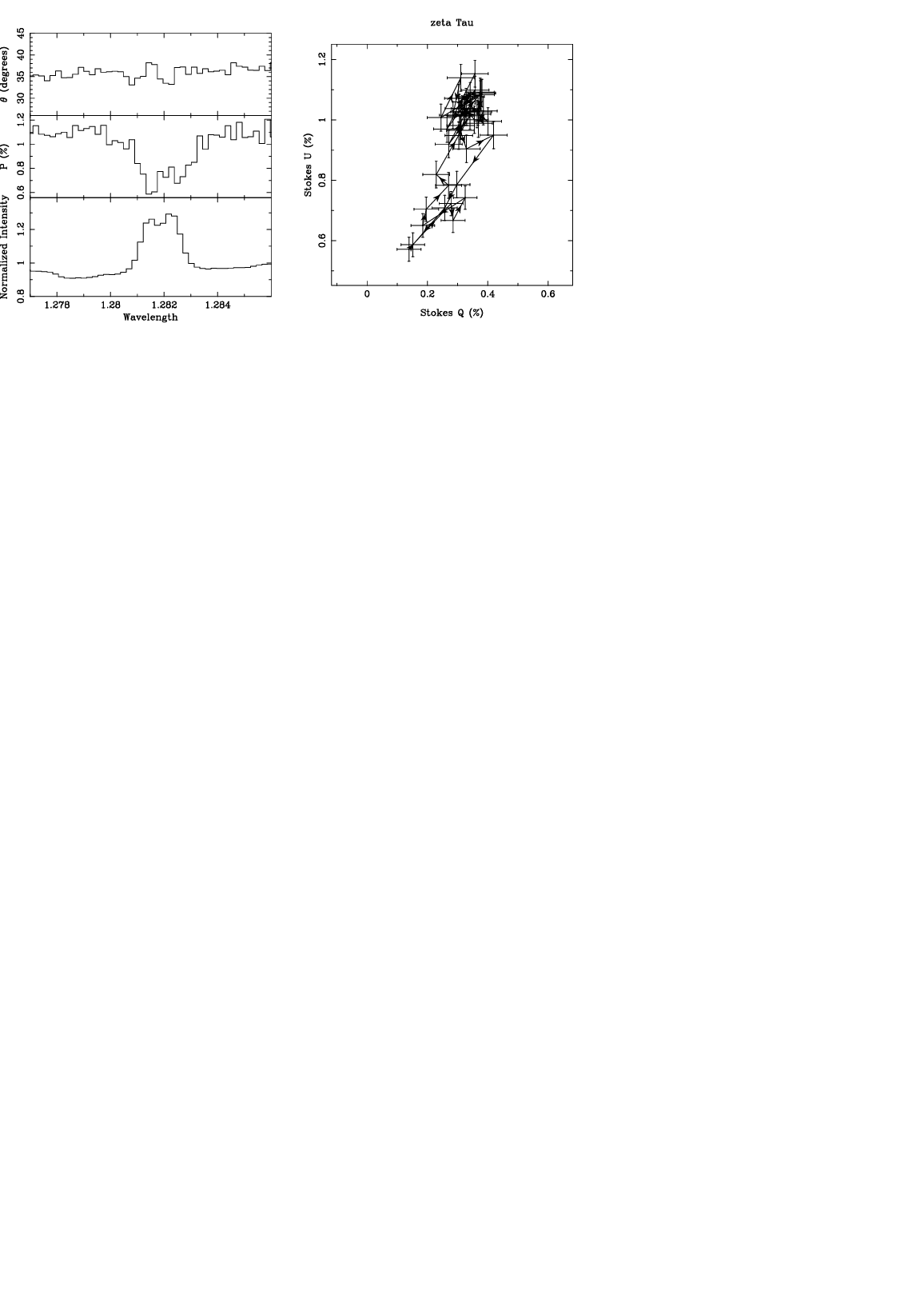}}
\caption{ The left hand plot shows the polarisation data of $\zeta$
Tau represented as a function of wavelength in $\mu$m, the bottom
panel shows the (continuum-normalised) intensity spectrum, while the
middle and upper panels show the polarisation and position angle
respectively. The latter two are rebinned to a corresponding accuracy
in the polarisation of 0.05\% . The right hand plot shows the Stokes
QU vectors with the same binning applied. The excursion of the
depolarisation in the QU plane implies a best fitting intrinsic PA of
32\degree.
\label{zeta} }
\end{figure*}

\section{Results}

Here we discuss the results individually for the objects.  We start by
describing the results of the classical Be star $\zeta$ Tau, and then
continue with the massive young stars.  In Table 1 the measured
polarisations of the objects are presented.

\subsection{  $\zeta$ Tau}

Because $\zeta$ Tau has a long history of (spectro)polarimetric
observations, it is an ideal object to test the diagnostic power of
near-infrared spectropolarimetry.  Also, $\zeta$ Tau is known to have
a large-scale disk found through interferometry, consistent with the
intrinsic line polarisation at optical wavelengths (Quirrenbach et
al. 1997).  Our data are presented in Fig.~\ref{zeta}, where the
Stokes I (intensity) vector is plotted in the bottom panel, and the
polarisation percentage and polarisation angle (PA) are displayed in
the middle and upper panels respectively.  The polarisation across the
Pa$\beta$ line shows a marked drop with respect to the continuum.

The continuum polarisation of 1.1\% is slightly lower and the PA is
very close to the optical values of Quirrenbach et al. (1997, their
mean values are about 1.5\% and 31\degree ). The small change in
polarization from the optical to the NIR is not expected from a
Serkowski law, where the dust polarisation drops off steeply with
wavelength, but is consistent with the flatter dependence from
electron scattering. We can measure the {\it intrinsic} PA from the
excursion across the line profile observed in the {\it QU} diagram;
$\Theta = \frac{1}{2}\times$ atan($\Delta$U/$\Delta$Q) yielding
32$\pm$1\degree .  As scattering in the optically thin case gives a PA
perpendicular to the disk, this compares very well with the disk's position
angle of$-58\pm4$\degree \ measured interferometrically by Quirrenbach
et al. (1997).

In summary, the near-infrared spectropolarimetry is a complementary
technique to optical spectropolarimetry and interferometry allowing us
to move to longer wavelength ranges where we can obtain detailed
observational constraints on types of object that would otherwise
remain obscured.

\subsection{MWC 349A} 

\begin{figure*}
 \mbox{\epsfxsize=0.\textwidth\epsfbox[1 575 350 775]{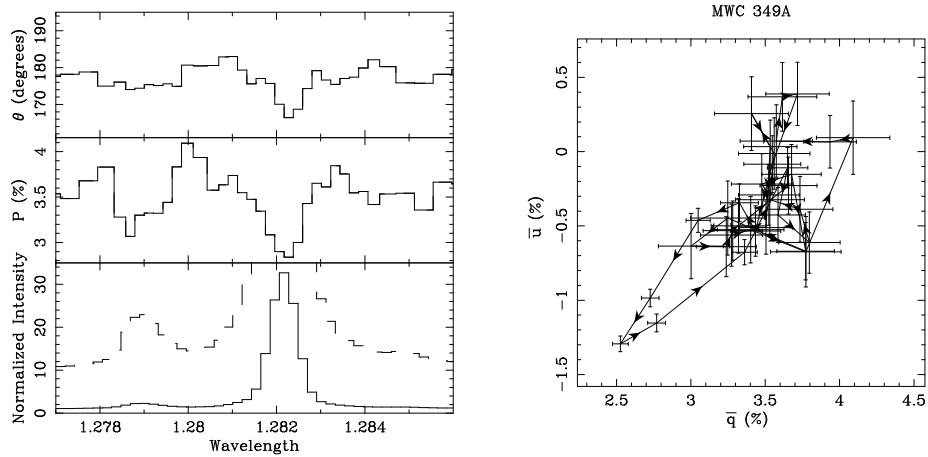}}
\caption{ As previous figure. The polarization data of MWC 349A
rebinned to 025\%.  The dashed line is an enhanced version of the
intensity spectrum intended to reveal the relatively strong, twice the
continuum, He {\sc i} $\lambda$ 1.27842$\mu$m line.  The polarization
spectrum shows an effect across both the Pa$\beta$ and He {\sc i}
emission lines. The depolarization is smaller across the He {\sc i}
line because of the relatively larger contribution of the continuum to
the observed polarization.
\label{m349} }
\end{figure*}

This well-studied object is mostly considered to be a massive pre-main
sequence star (e.g. Danchi, Tuthill \& Monnier 2001; Meyer, Nordsieck
\& Hoffman 2002) although it is sometimes proposed to be an evolved
massive star (strong proponents are Hofmann et al. 2002).  It had not
been previously observed by us at H$\alpha$ as the object is too faint
in the optical for medium resolution spectropolarimetry.
Our data are plotted in Fig.~\ref{m349}. The Pa$\beta$ line is
strongly in emission, whilst the He {\sc i} 1.27 \mic \ line, blueward
of Pa$\beta$, is clearly present as well.  The observed continuum
polarization of 3.5\%, at a PA of $\sim$ 180\degree, is consistent
with optical values (Yudin 1996). There is a clear signature of a line
effect at Pa$\beta$, particularly seen as a drop in polarization
percentage across the line (middle panel).

Representing the data in {\it QU} space, such as shown in
Fig.~\ref{m349} allows one to obtain the intrinsic PA of the
polarisation {\it independent} of any foreground polarization, which
only adds a constant {\it QU} vector to the intrinsic polarization.
There is a strong effect over the line, with a magnitude of $\sim$
1.5\% at a polarization angle of 21$\pm$1\degree \ as determined from
a weighted least squares fit through the {\it QU} data points, as above. A
similar, less strong effect at the same PA is visible for the helium
line. There is one earlier report of a `line-effect' in MWC 349A:
Meyer et al. (2002) obtained low resolution (7.5-10 \ang) optical
spectropolarimetry, and reported strong line depolarisations.
Earlier, broad-band optical polarimetry accompanied by narrow-band
\ha\ polarimetry was obtained by Zickgraf \& Schulte-Ladbeck (1989),
but they did not report the line-effect, possibly because it was
smeared out over the large wavelength range covered by the H$\alpha$
filter.

We also note that although the polarization angle changes by about
10\degree\ over the emission line, this does not necessarily have to
be an intrinsic property of the polarization.  One measures the vector
sum of both intrinsic and interstellar polarization. Due to this
vectorial nature, a simple depolarisation can change into a more
complicated profile with the addition of ISP (see e.g. Oudmaijer et
al. 1998 for a marked change in the case of HD 87643). The linear
excursion observed in {\it QU} space strongly suggests that we deal
with a straightforward depolarisation.

The obvious question that now arises is where does the scattering
arise, in the well-known ionised North-South bipolar outflow visible
in the radio (White \& Becker 1985) or the East-West circumstellar
disk as is evident from imaging of hydrogen recombination line masers
(e.g. Planesas, Martin-Pintado \& Serabyn 1992; Rodriguez \& Bastian
1994)?  As mentioned in the previous subsection, in the optically thin
case we will measure a polarization angle perpendicular to the disk
plane, while modelling of Be star disks implies polarization
percentages not much more than 2\% (Cassinelli et al. 1987; Waters \&
Marlborough 1992), with the polarization decreasing with inclination
angle.

The intrinsic position angle of the polarization is 21 $\pm$ 1\degree
, which is almost perpendicular to the position angle of
107$\pm$7\degree , measured from the H30$\alpha$ line maser (Planesas
et al. 1992) and the 100$\pm$3\degree measured from interferometric
images tracing warm dust (Danchi et al. 2001). This is consistent with
the scattering occurring in a structure at the same PA. As the larger
scale disk is observed to be close to edge-on, the depolarization of
about 1.5\% lends extra support to the idea that the
spectropolarimetry, and associated line effects, are due to a
circumstellar ionized disk.

Since MWC 349A is such a well-observed object, we now consider how our
Pa$\beta$ data fit in with existing ideas about the ionized gas around
this object.  Spectropolarimetry should be particularly helpful as it,
in principle, probes the ionized gas close to the star.

Hamann \& Simon (1986, 1988) proposed a model of the circumstellar
material, based on velocity resolved optical and near-infrared
spectroscopy of the star. They found that the line-width, and peak
separation of resolved hydrogen recombination lines decreased as
function of distance to the star. The Helium recombination lines
showed the largest widths, and were suggested to be originating at the
inner edge of a Keplerian rotating circumstellar disk at a distance
of $\simeq$ 2 AU from the star. The hydrogen recombination lines,
especially optically thick lines like H$\alpha$, were placed much
further from the star, consistent with the observed extent of the
hydrogen masers at 60 AU and more (e.g. Planesas et al. 1992). The
authors explicitly introduced a low density region between the star
and the disk/wind. This is mainly because of the observed line-widths:
if MWC 349 is indeed a massive hot star, the line widths of order 100
\kms\ are too small to be identified with a hot stellar wind whose
observed velocities are generally larger.  Instead, Hamann \& Simon
locate the ionised gas in a photo-evaporating disk (see also
Hollenbach et al. 1994), further away from the star, where the local
escape velocity is low.  This would present a coherent picture, where a
comparatively low density fully ionized inner disk is responsible for
the observed electron-scattering induced polarization.

We wish to mention a possible alternative that can explain the low
outflow velocity and still be consistent with a circumstellar disk.
Drew, Proga \& Stone (1998) { present a hydrodynamical  }model of
radiation driven disk winds for YSOs that also predict narrow H{\sc i}
line emission in the equatorial plane, and higher velocity, more
tenuous emission in the polar regions. Such a model could also be
responsible for the observed line-profiles in MWC 349A.

\subsection{MWC 342}

\begin{figure*}
\mbox{\epsfxsize=0.6\textwidth\epsfbox[1 600 300 800]{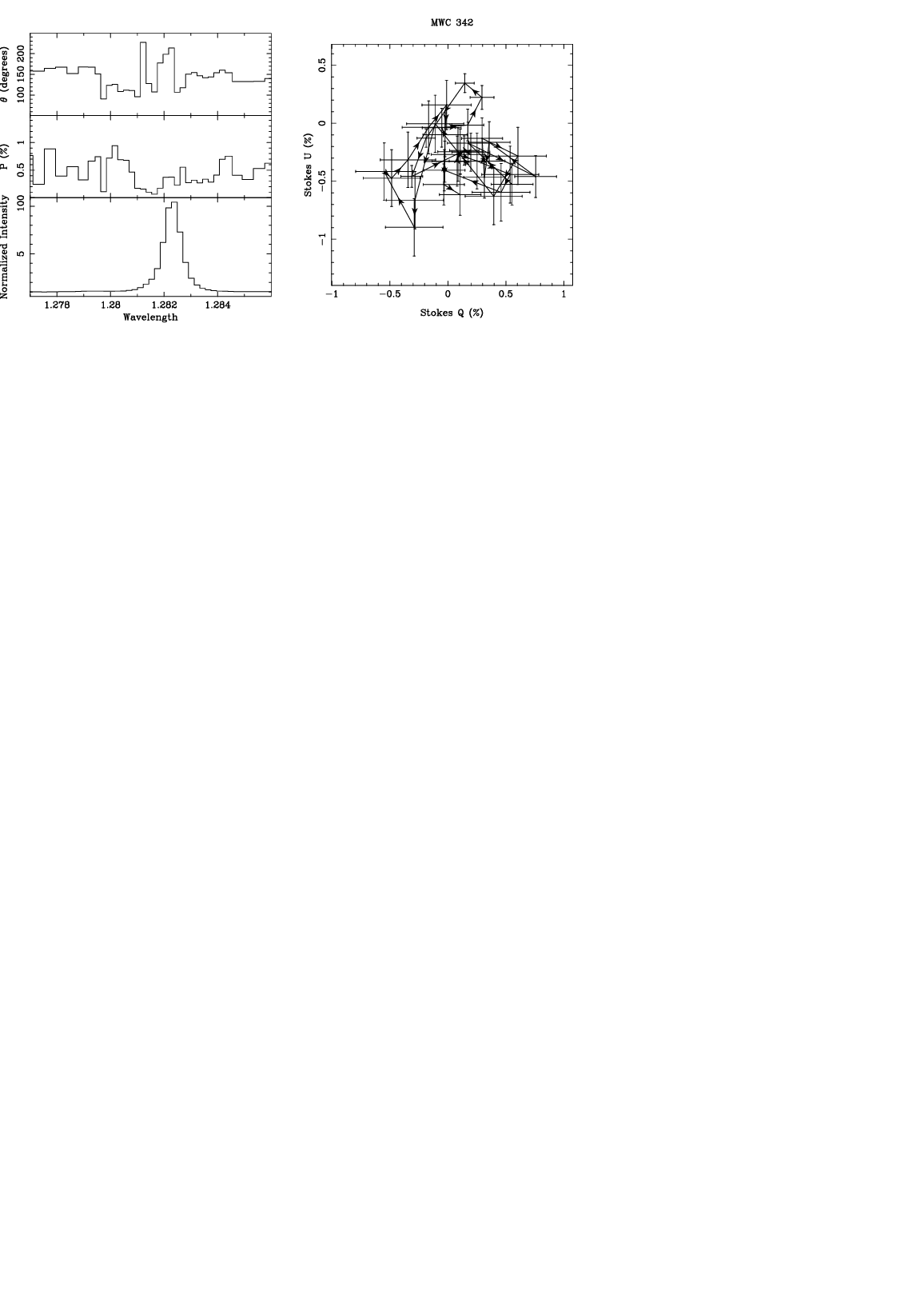}}
\caption{  As previous figures, but now the polarization data of MWC
342 rebinned to 0.25\% are plotted. The large errorbars combined with
the low polarization prevent the detection of  a line-effect.
\label{m342} }
\end{figure*}

This object, often classed a young Herbig Be star, has a rich emission
line spectrum with H$\alpha$ displaying a P Cygni profile, indicating
current outflow, while the presence of many forbidden emission lines
class the object a B[e] star (Jaschek \& Andrillat 1999).  The most
in-depth study to date of MWC 342 is probably that of Miroshnichenko
\& Corporon (1999) who suggest the object is a luminous evolved star
based on its extinction distance. The literature does not report any
observations that specifically study the geometry of the circumstellar
material. Clues that at least some asymmetries are present come from
the variable optical polarization reported by Bergner et al. (1990).
The continuum polarization around Pa$\beta$ of this object is 0.4\%
This makes it the least polarized of the objects in our sample and
corresponds to a phase of low optical polarization (Bergner et al
1990, Miroshnichenko, private communication).  { The data are
plotted in Fig.~\ref{m342} but no line effect can be inferred with
confidence. Since the polarization across the line measured per pixel
is only at the 2$\sigma$ level - as opposed to the value derived in
Table~\ref{targets} which is derived from many pixels - it is hard to
measure an effect at a significant level, and easy to overinterpret
such data. Rebinning the data would yield severely undersampled data
and potentially lead to misleading results. } The uncertainty in
polarization of the individual pixels is 0.25\%, so the upper limit to
the line-depolarisation is at least 0.75\%, as an entire resolution
element needs to be covered.

\subsection{GL 490}

AFGL 490 is a heavily embedded massive young star with a luminosity of
order 10$^3$L$_{\odot}$ (e.g. Schreyer et al. 2002). It is subject to
an optical extinction $>$ 20 (Bunn et al. 1995), and is surrounded by
a thick dusty disk at a PA of around 120-145 \degree . The disk
structure has been observed using a variety of techniques:
near-infrared imaging was done by Minchin et al. (1991); high
resolution molecular CS observations were taken by Schreyer et
al. (2002), whilst Alvarez et al. (2004) detect a, possibly transient,
structure in near-infrared speckle interferometry. In addition, a
flattened radio structure was found by Campbell, Persson \& McGregor
(1986).  The central object powers a wide bi-polar CO outflow roughly
perpendicular to the disk (e.g. Mitchell et al. 1995, Schreyer et
al. 2002).

{\ The optical polarization was studied by Haas, Leinert \& Lenzen
  (1992). They found that the bulk of the optical polarization of
  $\sim$18\% can be explained by a combination of foreground
  polarization (surrounding stars have optical polarizations of up to
  13\%) and polarization by circumstellar dust. As electron scattering
  typically would contribute around 1-2\%, it is indeed most likely
  the dust scattering, rather than electron scattering that
  contributes most to the circumstellar polarization.  Because the
  dust can not exist close to the star owing to its comparatively low
  condensation temperature, the Pa$\beta$ line forming region and the
  central source continuum will be seen as a point sources by the dust
  and be equally polarized.  Although the observed polarization is a
  combination of interstellar dust and circumstellar dust and
  electron-scattering, the ``intrinsic'' electron scattered radiation
  is revealed in {\it QU} space, as the interstellar and circumstellar
  dust polarization will add respective constant {\it QU} polarization
  vectors, whereas the intrinsic electron scattering polarization can
  be dependent on the velocities.}

\begin{figure*}
 \mbox{\epsfxsize=0.6\textwidth\epsfbox[1 600 300 800]{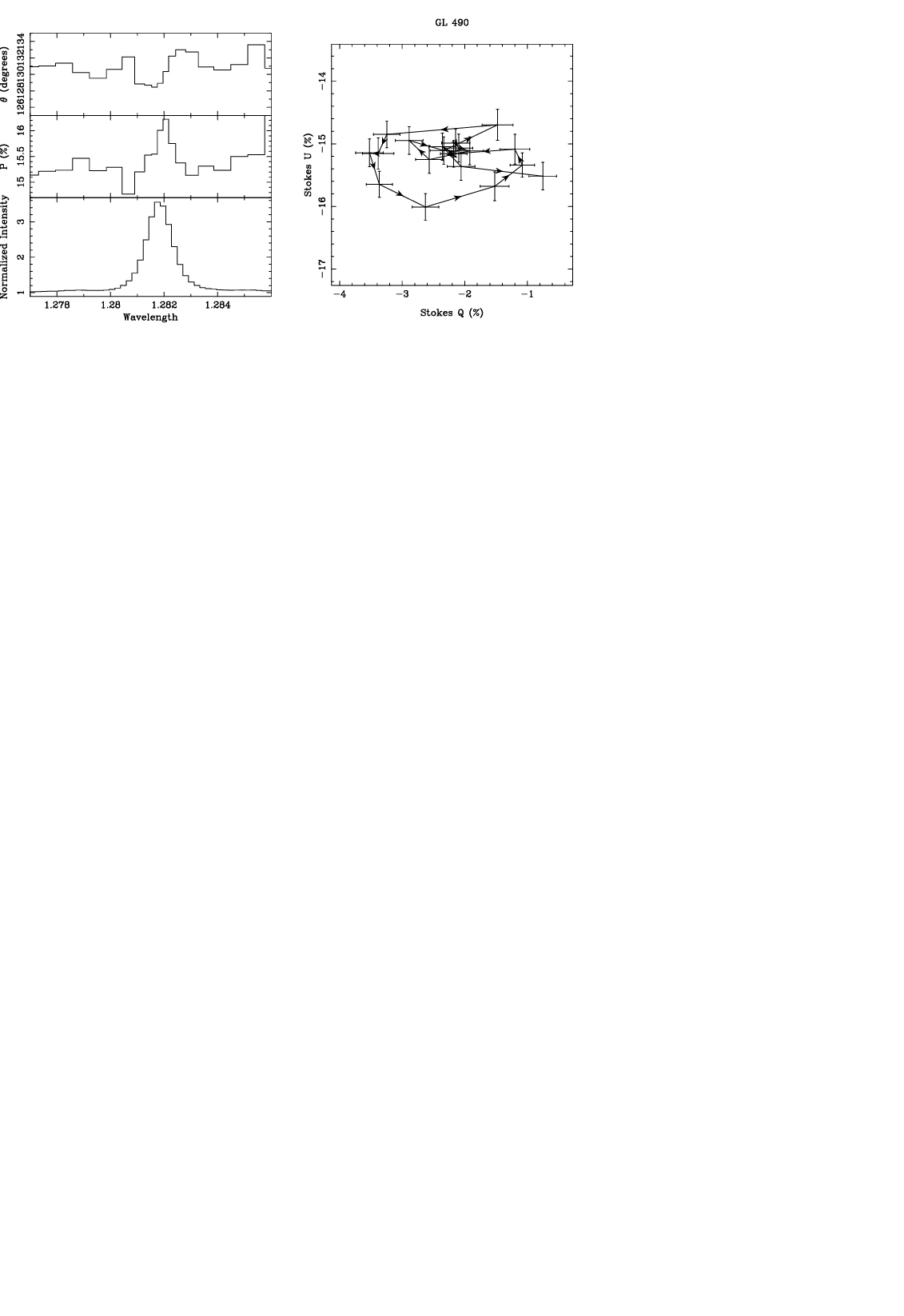}}
\caption{ As the previous figures. The polarization data of GL 490, 
rebinned to 0.25\% per pixel.
\label{g490} }
\end{figure*}

Our polarimetric spectrum of GL 490 is presented in
Fig.~\ref{g490}. The emission line itself is resolved as it is
slightly wider than the instrumental profile, thereby allowing the
study of changes in the polarization characteristics across the line
itself.  We measure a continuum polarization of $\sim$ 15\% at a PA of
130\degree, which is consistent with the PA measured by Haas et
al. (1992) in optical continuum polarization measurements, and also
with with the large scale dusty disk seen using imaging polarimetry
(Minchin et al. 1991).  In the polarisation triplot, there appears
evidence of changes in line polarisation, in particular a rotation in
the PA.  This is most likely a real feature because the emission line
is resolved in our data while the feature is larger than the formal
errorbars in both the PA spectrum, and the {\it QU} graph.

We note that this effect may well be similar to the line effects
measured in many Herbig Ae stars by Vink et al. (2002). In their work
on these objects they found that the polarisation changes across
H$\alpha$ were narrower than the width of the intensity profile itself
-- which is {\it}inconsistent with depolarisation.  Instead, the data
showed line polarisations that are thought to be due to \ha\ emission
originating from a compact source of photons that scatters off an
exterior rotating medium. In contrast to the effect of depolarisation
observed for MWC 349A, these line polarisations may yield kinematic
information.  The PA rotations across the line profile translate into
``loops'' when the same data are plotted in $QU$ space.


 Indeed, the excursion in QU space resembles a loop, rather than the
straight line observed for MWC 349A. Comparison with models by Vink et
al (2005a, see also Wood, Brown \& Fox 1993) indicates that such
behaviour is the result of the presence of a rotating disk { with
an inner hole.  For comparison, rotating disks without an inner hole,
extending down to the stellar photosphere would display a double loop
in the QU diagram. }If the detection is can be confirmed, it provides
strong clues that GL 490 is surrounded by a disk and is still actively
accreting material.

\section{Summary}

We have presented the first hydrogen recombination line
spectropolarimetric observations in the near-infrared.  The results
from this  study are encouraging. Our targets were two
optically bright Herbig Be type objects (MWC 349A and MWC 342), an
optically faint massive Young Stellar Object, GL 490 and for
comparison the well-known classical Be star $\zeta$ Tau, known to be
surrounded by an ionised disk.

Three of these stars show a line effect. For $\zeta$ Tau and MWC 349A
the line-effect is explained as being due to optically-thin electron
scattering in a circumstellar disk. These objects have high-resolution
imaging data available and in both cases the position angle of the
(larger scale) disk on the sky is consistent with the angle derived
from the polarization data.  This validates spectropolarimetry as a
good means to detect disks on small scales.

The preliminary detection of a PA rotation in GL 490 may indicate the
presence of a small scale rotating accretion disk { with an inner
hole} similar to those recently discovered in H$\alpha$ in Herbig Ae
and T Tauri stars (Vink et al. 2002, 2005b).

\smallskip

\noindent
{\it Acknowledgements :} We thank Tim Harries for discussing the
intricacies of reducing near-infrared spectropolarimetric data. We
thank Anatoly Miroshnichenko for providing us with machine readable
data from the Bergner et al. (1990) paper, supplemented by additional
unpublished polarimetry.  The United Kingdom Infrared Telescope is
operated by the Joint Astronomy Centre on behalf of the UK Particle
Physics and Astronomy Research Council. JSV is funded by PPARC.  IRAF
is written and supported by the IRAF programming group at the National
Optical Astronomy Observatories (NOAO) in Tucson, Arizona. NOAO is
operated by the Association of Universities for Research in Astronomy
(AURA), Inc. under cooperative agreement with the National Science
Foundation

\end{document}